\documentclass[twocolumn,prl,showpacs,amssymb,superscriptaddress,amsmath]{revtex4}

\usepackage{bm}
\usepackage{graphicx}
\usepackage{color}
\usepackage{dcolumn}

\newcommand{\vK}{\mbox{$\vec K $}}
\newcommand{\vk}{\mbox{$\vec k $}}

\begin{document}

\title{Recoil Effect of Photoelectrons in the Fermi-Edge of Simple Metals}

\author{Y. Takata}
\email{E-mail: takatay@spring8.or.jp}
\affiliation{RIKEN SPring-8 Center, Sayo-cho, Hyogo 679-5148, Japan}
\author{Y. Kayanuma}
\email{E-mail: kayanuma@pe.osakafu-u.ac.jp}
\affiliation{Graduate School of Engineering, Osaka Prefecture University, Sakai, Osaka, 599-8531 Japan}
\author{S. Oshima}
\affiliation{Graduate School of Engineering, Osaka Prefecture University, Sakai, Osaka, 599-8531 Japan}
\author{S. Tanaka}
\affiliation{Graduate School of Sciences, Osaka Prefecture University, Sakai, Osaka, 599-8531, Japan}
\author{M. Yabashi}
\affiliation{RIKEN SPring-8 Center, Sayo-cho, Hyogo 679-5148, Japan}
\affiliation{JASRI, SPring-8, Sayo-cho, Hyogo 679-5198, Japan}
\author{K. Tamasaku}
\affiliation{RIKEN SPring-8 Center, Sayo-cho, Hyogo 679-5148, Japan}
\author{Y. Nishino}
\affiliation{RIKEN SPring-8 Center, Sayo-cho, Hyogo 679-5148, Japan}
\author{M. Matsunami}
\affiliation{RIKEN SPring-8 Center, Sayo-cho, Hyogo 679-5148, Japan}
\author{R. Eguchi}
\affiliation{RIKEN SPring-8 Center, Sayo-cho, Hyogo 679-5148, Japan}
\author{A. Chainani}
\affiliation{RIKEN SPring-8 Center, Sayo-cho, Hyogo 679-5148, Japan}
\author{M. Oura}
\affiliation{RIKEN SPring-8 Center, Sayo-cho, Hyogo 679-5148, Japan}
\author{T. Takeuchi}
\affiliation{RIKEN SPring-8 Center, Sayo-cho, Hyogo 679-5148, Japan}
\author{Y. Senba}
\affiliation{JASRI, SPring-8, Sayo-cho, Hyogo 679-5198, Japan}
\author{H. Ohashi}
\affiliation{JASRI, SPring-8, Sayo-cho, Hyogo 679-5198, Japan}
\author{S. Shin}
\affiliation{RIKEN SPring-8 Center, Sayo-cho, Hyogo 679-5148, Japan}
\affiliation{Institute for Solid State Physics, University of Tokyo, Kashiwa, Chiba 277-8581, Japan}
\author{T. Ishikawa}
\affiliation{RIKEN SPring-8 Center, Sayo-cho, Hyogo 679-5148, Japan}
\affiliation{JASRI, SPring-8, Sayo-cho, Hyogo 679-5198, Japan}

\date{\today}

\begin{abstract}
\textbf{ABSTRACT}
High energy resolution photoelectron spectroscopy of conduction electrons in the vicinity
of the Fermi-edge in Al and Au at the 
excitation energy of 880 and 7940eV was carried out using synchrotron radiation. For the excitation energy of 7940eV, the observed Fermi energy of Al shows 
a remarkable shift to higher binding energy as compared with that of Au, with accompanying broadening. This is due to the recoil effect 
of the emitted photoelectrons. The observed spectra are well reproduced by a simple model of Bloch electrons based on the isotropic Debye model.
\end{abstract}
\maketitle

In x-ray photoelectron spectroscopy (XPS), the emitted electron kicks the atom from which it is ejected in accordance with the conservation of 
momentum\cite{Flynn,Domcke,Cini}. This gives rise to a loss of the kinetic energy of the emitted photoelectron. This effect is usually negligible because of the enormous 
mass difference between the atom and the electron, as long as the excitation energy is not very large.@Quite recently, however, a clear evidence of 
recoil effects has been found in the hard x-ray photoelectron spectra of graphite\cite{Takata}. Under the excitation of the core level by x-ray photons 
with energy of several keV, the photoelectron spectra show a remarkable shift and broadening as compared 
to the case of excitation by soft x-rays. Recoil effects of carbon $1s$ photoelectrons have also been reported 
in CH$_4$ molecules\cite{Kukk} and CF$_4$ molecules\cite{Thomas}, in this case as a recoil induced modification of vibrational structures.
\par
Based on a simple picture of an atom at rest in vacuum, the recoil energy $\Delta E$ imparted to the atom with mass $M$ by 
a photoelectron with mass $m$ and kinetic energy $E_K$ is estimated as
\begin{equation}
\Delta E=E_K \times (m/M).
\end{equation}
This recoil energy $\Delta E$ is observed as an apparent increase of the binding energy of the core electron. In solids, 
the recoil energy is absorbed by the phonon bath, resulting in the excitation of phonons. 
Actually, the observed photoelectron spectra for C $1s$ in graphite have been well reproduced by an 
anisotropic Debye model\cite{Takata}, which 
takes into account solid state effects appropriate to graphite\cite{Wirtz}. The recoil effects of the core electrons are characterized 
by the peak shift and the asymmetric broadening depending on the incident x-ray energy. Subsequently, such characteristic features have 
been observed in the core level XPS not only for $1s$  of graphite but also in other materials such as Be $1s$ in Be metal, B $1s$
in MgB$_{2}$ and $2p$ level in Al metal\cite{Takata2}. The existence of remarkable recoil effects in the hard x-ray photoelectron spectra is now well established as far as the core level is concerned.
\par
Similar spectroscopic features of recoil effects have been observed in other fundamental processes such as elastic electron 
scattering\cite{Vos1,Went} and neutron scattering\cite{Rauh,Fielding}. These spectra can be understood essentially by 
the same principle of momentum conservation\cite{Vos2}. In contrast to the electron scattering and the neutron scattering, 
however, the photoelectron spectra tell us information on the specific electronic state from which the electron is ejected. 
A natural question then arises: Are there any recoil effects in the photoelectron spectra for valence levels? 
Since the Bloch electrons in the valence bands are delocalized all over the crystal, it may be considered 
at first sight that the recoil momentum is shared by all the atoms of the crystal 
so that there would be no observable recoil effect, just like the recoilless
transition in the M\"ossbauer effect\cite{Mossbauer}. In the present Letter, we show, 
for the first time, a clear evidence of the recoil effect 
for conduction electrons in a simple metal. The experimental data for XPS in the vicinity of the Fermi-edge of Al 
indicate a remarkable shift and broadening 
depending on the excitation energy. 
The observed spectra are well reproduced by a theory which takes into account the momentum conservation of the 
Bloch electrons expanded in a Wannier function basis. 

High resolution photoelectron spectroscopy of conduction electrons in the vicinity of the Fermi-edge was performed at SPring-8 using 
synchrotron radiation. Hard x-ray spectra at the excitation energy of 7940 eV and soft x-ray spectra of 880eV
were measured at the undulator beamlines BL29XU~\cite{Tamasaku, Ishikawa} and BL17SU~\cite{Ohashi}, respectively, 
using hemispherical electron energy analyzers. 
The thick films of Au and Al on Cu substrates were prepared by evaporation in the UHV preparation chamber, and 
were directly transferred into the measurement chamber. No signal from the substrate Cu or 
surface contamination was observed in both hard and soft x-ray photoelectron spectra.
The energy scale of the spectra were calibrated very accurately ($<$5 meV) by fitting the Fermi-edge spectra of Au. 

Figure 1 shows the photoelectron spectra around the Fermi-edge of Au (squares) and Al (circles) measured at 20K with the 
excitation energy of 7940 eV. The total instrumental energy resolution for both the spectra, as determined by the 
the beamline crystal monochromator and the electron energy analyzer,
is the same. The binding energy scale is calibrated by assuming that the recoil energy in Au (M=197) is negligible.
It is clear that the Fermi-edge of Al (M=27) is shifted to higher binding energy. This shift is due to the kinetic energy loss of 
photoelectron from the Fermi level in Al because of its lighter 
atomic mass compared to Au, and is an evidence of the recoil effect of Bloch electrons.   
Furthermore, the edge profile has a slightly larger slope for Al than for Au. Conventional fitting analysis of these Fermi-edge profiles using the
Fermi-Dirac function  (not shown) elucidates an energy shift of 120 meV in Al relative to Au, and Gaussian widths of 160 meV for Al 
and 124 meV for Au. 
The broadening of the width in Al is also a sign of the recoil effect, because the contribution of the instrumental energy 
resolution to the width is the same in these spectra.  

Soft x-ray spectra of Au (squares) and Al (circles) measured at 50K with the 
excitation energy of 880 eV are shown in Fig. 2. In the wide range spectra (Fig. 2(a)), it is difficult to recognize the difference 
between the Au and Al. However, in the expanded spectra (Fig. 2(b)), the energy shift and the broadening of Fermi-edge profile 
is certainly observed. A fitting analysis clarifies a energy shift of 12 meV in Al relative to Au, and Gaussian width of 140 meV in Al 
and 118 meV in Au.   

\begin{figure}
\centering
\includegraphics[width=7cm]{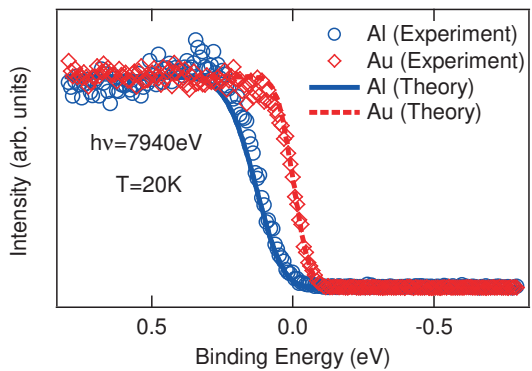}
\caption{(color online) Photoelectron spectra near Fermi-edge of Al (circles) and Au (squares) with excitation energy 7940eV. 
The zero point of the binding energy is chosen at the observed value of the chemical potential of Au. The solid line and the 
dotted line indicate the theoretical curves calculated by the Debye model.}

\label{tfig:1}     
\end{figure}

Consider the transition probabilty $I(\vk,\vK)$ in which a Bloch electron with 
wave vector $\vk$ is emitted to the free electron state with wave vector $\vK$ by an x-ray photon with energy $h\nu$. 
We neglect the momentum of the incident photon since it is an order of magnitude smaller than that of the emitted electron 
in this energy region. Without recoil effect, the component of the wave vector parallel to the surface is conserved in the periodic zone scheme. The perpendicular component is not conserved, but is determined by the conservation of energy. @In the presence of recoil effect, it is written as
\begin{equation*}
\epsilon_k+h\nu + \epsilon_m=E_K+\epsilon_n,
\end{equation*}
in which $\epsilon_k$ is the energy of the Bloch electron measured from the vacuum level 
and $\epsilon_m$ and $\epsilon_n$ are the energies of 
the lattice vibrations in the initial state and the final state, respectively. 
The interaction Hamiltonian with the x-ray photon is given by 
$H_I=\left(a+a^\dagger\right)\vec \kappa\cdot \vec p$, aside from irrelevant factors, where $a$ is the annihilation operator for the 
x-ray and $\vec p$ is the momentum of the electron, and $\vec \kappa$ is the polarization vector of the photon.
\par
The initial state of the transition is given by 
$|\Psi_i\rangle=|h\nu\rangle\bigotimes |\psi_k\rangle\bigotimes |m\rangle$, where $|h\nu\rangle$ is 
the one photon state, $|\psi_k\rangle$ is the Bloch state 
(we suprress the band index here), and $|m\rangle$ is a phonon state. 
The Bloch state $\langle \vec r|\psi_k\rangle\equiv \psi_k(\vec r)$ is expanded by the Wannier functions as
\begin{equation}
\psi_k(\vec r)=N^{-1/2}\sum_i e^{i\vec k\cdot \vec R_i}w(\vec r -\vec R_i),
\end{equation}
where $\vec R_i$ is the lattice vector and $N$ is the number of atoms. In the above equation, $\vec R_i$ is usually regarded as 
a parameter fixed at the equilibrium point of the lattice $\vec R_i^0$. In order to describe the recoil effect, 
we consider it as a dynamical variable fluctuating 
around $\vec R_i^0$ as $\vec R_i=\vec R_i^0 +\vec u_i$, where $\vec u_i$ is the displacement vector. 
The Wannier function is assumed to follow 
this lattice displacement adiabatically. The lattice fluctuation gives rise to the local 
change of the band energy, which is nothing but 
the electron-phonon interaction represented by
the deformation potential interaction. We neglect here the electron-lattice interaction for simplicity. 
The final state of the transition is given by $|\Psi_f\rangle=|0 \rangle\bigotimes |\varphi_K\rangle\bigotimes |n\rangle$,
in which $|0\rangle$ is the vacuum of the photon, $|\varphi_K\rangle$ is the plane wave of the electron 
$\langle \vec r|\varphi_K\rangle=\left(2\pi\right)^{-3/2}\exp\left(i\vK\cdot\vec r\right)$ with energy 
$E_K=\hbar^2\vK^2/2m$ and $|n\rangle$ is a phonon state of the crystal. 
\par

\begin{figure}
\centering
\includegraphics[width=7cm]{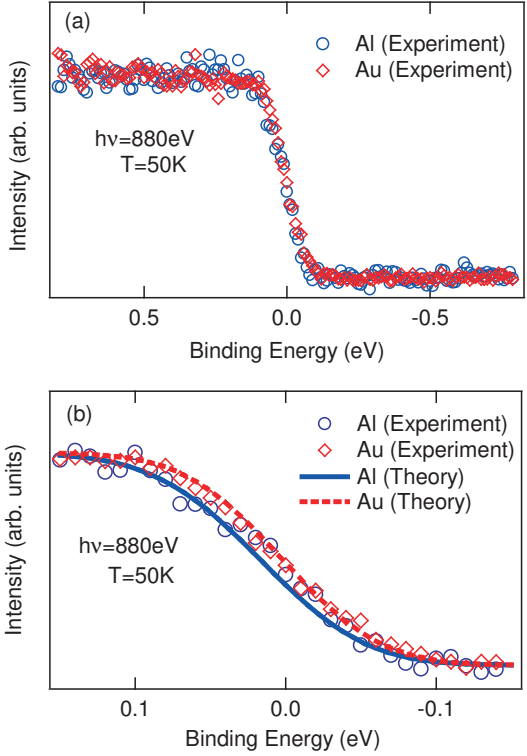}
\caption{(color online) (a)Photoelectron spectra near Fermi-edge of Al (circles) and Au (squares) with excitation energy 880eV. 
The zero point of the binding energy is chosen at the observed value of the chemical potential of Au. (b)Expanded scale plots of 
the spectra. The solid line and the 
dotted line indicate the theoretical curves calculated by the Debye model.}
\label{tfig:2}     
\end{figure}

By Fermi's golden rule, we have
\begin{eqnarray}
I\left(\vk,\vK\right)&=&\Bigl< \sum_f|\langle\Psi_f|H_I|\Psi_i\rangle|^2 \nonumber \\
                   &\times& \delta(E_K+\epsilon_n-h\nu-\epsilon_k-\epsilon_m)\Bigr>,
\end{eqnarray}
in which the summation $\sum_f$ runs over the final phonon states and $<\cdots>$ means the canonical average over 
the initial phonon states. By using the representation of the Bloch function (1), the transition matrix element is 
given by
\begin{equation}
\langle\varphi_K|H_I|\psi_k\rangle=\vec \zeta\cdot\vec\epsilon \frac{1}{\sqrt{N}}\sum_ie^{-i(\vK-\vk)\cdot\vec R_i^0}e^{-i\vec K\cdot\vec u_i},
\end{equation}
where we have changed the integration variable from $\vec r$ to $\vec r-\vec R_i$ in the evaluation of $i$th term, and set
\begin{equation*}
\vec{\zeta} = (2\pi)^{-3/2}\int\!d^3 r e^{-i\vK\cdot\vec{r}} \left( -i\hbar \frac{\partial}{\partial \vec{r}} \right) w(\vec r).
\end{equation*}
Putting Eq.(4) into (3), and using the translational symmetry, we find after some exercise,
\begin{eqnarray}
I(\vec k,\vec K)&=&\frac{|\vec \zeta\cdot \vec \kappa|^2}{2\pi\hbar}\int_{-\infty}^\infty dt e^{-i(E_K-h\nu-\epsilon_k)t/\hbar}\nonumber \\
&\times&\sum_ie^{i(\vK-\vk)\cdot\vec R_i^0}F_i(t)
\end{eqnarray}
with the spatiotemporal generating function
\begin{equation*}
F_i(t)=\left<e^{i\vK\cdot\vec u_i(t)}e^{-i\vK\cdot\vec u_0}\right>,
\end{equation*}
where $\vec u_i(t)$ is the Heisenberg representation of $\vec u_i$ at time $t$, and we have chosen $\vec R_0^0=0$. 
The generating function can be calculated by expanding $\vec u_i$ with the phonon operators as in Ref.[3],
\begin{equation*}
F_i(t)=\exp\left[\int_{-\infty}^\infty\left\{J_i(\omega)e^{-i\omega t}-J_0(\omega)\right\}d\omega\right]
\end{equation*}
in which
the spectral function $J_i(\omega)$ is given by
\begin{eqnarray*}
J_i(\omega)&=&\sum_q \alpha_q^2\big[\left\{n(\omega_q)+1\right\}e^{i\vec q\cdot\vec R_i^0}
\delta(\omega-\omega_q) \nonumber \\
&+&n(\omega_q)e^{-i\vec q\cdot\vec R_i^0}\delta(\omega+\omega_q)\big],
\end{eqnarray*}
with
\begin{equation*}
\alpha_q^2 = \left(
        \frac{\hbar}{2NM\omega_q} 
\right)\left|
                \vK \cdot \vec{\eta_q}
        \right|^2,
\end{equation*}
and
\begin{equation*}
n(\omega_q) = 1/\left( e^{\hbar \omega_q/k_B T} -1 \right),
\end{equation*}
where $q$ is the abbreviation for the wave vector and the branch index of phonons, and $\vec \eta_q$ is the polarization vector of the phonon.
\par

The actual calculation was done by assuming the Debye model for the phonons. The spectral function is given by
\begin{eqnarray}
J_i(\omega)=\Biggl\{   \Biggr.
\begin{array}{cc}
\frac{3g}{ \omega_D^2} \{ n(\omega)+1\} {c\over R_i^0}  
\sin(\frac{\omega R_i^0 }{c}) \;, & (0 < \omega < \omega_D) \\
\frac{3g}{ \omega_D^2} n(|\omega |) {c\over R_i^0} \sin(\frac{\omega  
R_i^0 }{c}) \;, & ( -\omega_D< \omega < 0 ) \\
\end{array}
\end{eqnarray}

in which $g\equiv E_K/\hbar\omega_D$ is the effective coupling constant, $\omega_D$ is the Debye frequency and $c$ is the sound velocity.
\par
In the absence of the recoil effect $F_i(t)=1$, so that Eq.(5) results in the conservation of the wave vector modulus 
a reciprocal lattice vector. The recoil effect, however, destroys the spatial coherence. The relative magnitude of the 
contribution to the total intensity of the term $i\neq 0$ to that $i=0$ is estimated much smaller than unity  
even for the terms corresponding to the nearest neighbor atoms in the region of hard x-ray. 
We can thus safely neglect those terms as $i\neq 0$ in accordance with the incoherent scattering approximation\cite{VanHove}. 
The angular dependence 
both for $\vk$ and $\vK$ disappears in the photoelectron spectrum. By introducing the binding energy $\epsilon$ for the initial state 
$\epsilon=-\epsilon_k-\phi$ with $\phi$ being the work function, and the apparent binding energy $E$ defined by 
$E=h\nu-E_K-\phi$, the photemission spectrum is written as 
\begin{equation*}
I(\epsilon,E)=\frac{1}{2\pi\hbar}\int_{-\infty}^\infty dt e^{i(E-\epsilon)t/\hbar}F_0(t).
\end{equation*}
where the angular dependence of the emitted electrons is averaged in accordance with the experiment. 
The actual spectrum is given by taking the average over the initial distribution of the electrons, 
\begin{equation*}
I(E)=\int I(\epsilon,E)f(\mu-\epsilon)D(\epsilon)d\epsilon,
\end{equation*}
where $f(\epsilon)=1/\left(e^{(\epsilon-\mu)/k_BT}+1\right)$ is the Fermi distribution function 
with the chemical potential $\mu$, 
and $D(\epsilon)$ is the density of state near the Fermi-edge which is approximated constant. 
Finally, we take into account the Gaussian broadening due to the resolution of the apparatus. 
\par
The theoretical curves for the XPS with excitation energy 7940eV are plotted in Fig.1 for Au (dotted line) 
and Al (solid line). 
The Debye energies are $\hbar\omega_D$(Al)=36.8meV and $\hbar\omega_D$(Au)=14.2meV\cite{Kittel}. The boadening of the 
observed Fermi-edge profile originates from three factors; the temperature dependence of 
the Fermi-Dirac function, the experimental 
resolution, and the recoil effect. The broadening due to the 
experimental resolution is fixed to be 108meV (FWHM) by fitting the line shape for Au.
The effective coupling constant for Al and Au are 
$g=4.3$ and $1.5$, respectively. In this figure, the zero 
of the binding energy is chosen at the experimental value of the chemical potential of Au. 
As shown in Fig.1, the agreement with the theoretical curves is good. The observed Fermi energy of Al 
is shifted by 120meV as compared with that of Au due to the recoil effect. A simple estimation by using Eq.(1) 
gives the value of this shift to be 138meV. The quantum mechanical calculation based on the Debye model correctly 
reproduces the experimental value. The reduction from 138meV to 120meV 
is due to the quantum effect of phonons\cite{Takata,Mossbauer}. 
It should be noted that, in the case of hard x-ray excitation with energy as high as 7940eV, 
even the observed Fermi energy of Au is shifted about 24meV from the true value. 
\par
In Fig.2 (b), the results for the case with excitation energy 880eV are shown. 
The effective coupling constants for Al and Au are $g=0.48$ and $0.17$, respectively, in this case. Although the difference of 
the spectrum between Al and Au is small in this excitation energy, it does exist.
\par
In this Letter, we have reported the modification of the photoelectron spectra at the Fermi-edge 
due to the recoil effect in simple metals. 
The existence of the photoelectron recoil effect means that the electron is coupled with the 
crystal lattice, and the wave function of the electron follows adiabatically the atomic motion. 
Therefore, it is a little surprising that the Bloch electron in the valence band of Al, which is a 
typical material where the nearly free-electron picture works well\cite{Harrison}, shows remarkable recoil effects. 
As shown above, Bloch's theorem itself guarantees the dependence of the wave function on the 
lattice coordinates. The recoil effect directly follows from this fact as a kinematic effect. The finding of recoil effects in the Bloch 
electrons indicates a new spectroscopic aspect in the XPS, and due care must be taken to interpret
changes at and near the Fermi level when using hard x-rays.   
The magnitude of the recoil effect depends essentially on the mass of the 
component atoms, as shown in Fig.1. Since the photoelctron spectra reflect 
also the nature of the electronic state in the initial state, the XPS recoil effect in the valence levels 
of \textit{composite materials} poses an interesting problem. If the material is composed of atoms 
with big mass differences, and if the valence levels are made of hybridized orbitals originating in specific atomic 
species, the modification of the photoelectron spectra will depend on the local density of state of the 
level. The experimental and the theoretical study of such effects is left for future works.


This work was partially supported by the Grant-in-Aid for Scientific Research (B) from Japan Society for 
the Promotion of Science (No. 20340079).  


\end{document}